\documentclass[nofootinbib,prd,aps,superscriptaddress,preprintnumbers,twocolumn,showpacs]{revtex4-1}

\usepackage[dvipdfmx]{graphicx}
\usepackage{amsmath,amssymb,amsthm,bm,bigdelim,multirow}
\usepackage{color}
\usepackage{booktabs}
\usepackage{braket,cases,latexsym}
\usepackage{here}

\begin{document}

\title{Phase transition of four-dimensional lattice $\phi^4$ theory with tensor renormalization group}

\author{Shinichiro Akiyama}
\email[]{akiyama@het.ph.tsukuba.ac.jp}
\affiliation{Graduate School of Pure and Applied Sciences, University of Tsukuba, Tsukuba, Ibaraki
    305-8571, Japan}

\author{Yoshinobu Kuramashi}
\email[]{kuramasi@het.ph.tsukuba.ac.jp}
\affiliation{Center for Computational Sciences, University of Tsukuba, Tsukuba, Ibaraki
    305-8577, Japan}

\author{Yusuke Yoshimura}
\email[]{yoshimur@ccs.tsukuba.ac.jp}
\affiliation{Center for Computational Sciences, University of Tsukuba, Tsukuba, Ibaraki
    305-8577, Japan}

\begin{abstract}
  We investigate the phase transition of the four-dimensional single-component $\phi^4$ theory on the lattice using the tensor renormalization group method.  We have examined the hopping parameter dependence of the bond energy and the vacuum condensation of the scalar field $\braket{\phi}$ at a finite quartic coupling $\lambda$ on large volumes up to $V=1024^4$ in order to detect the spontaneous breaking of the $\mathbb{Z}_2$ symmetry. Our results show that the system undergoes the weak first-order phase transition at a certain critical value of the hopping parameter. We also make a comparative study of the three-dimensional $\phi^4$ theory and find that the properties of the phase transition are consistent with the universality class of the three-dimensional Ising model. 
\end{abstract}

\date{\today}

\preprint{UTHEP-755, UTCCS-P-136}

\maketitle

\section{Introduction}
\label{sec:intro}

The issue of the triviality of the four-dimensional ($4d$) $\phi^4$ theory has been a theoretical concern among particle physicists, because it is related to the scalar sector in the standard model \cite{Wilson:1973jj,Aizenman:1981zz,Aizenman:1982ze,Frohlich:1982tw,Dashen:1983ts,Lindner:1985uk,Hasenfratz:1987eh,Luscher:1987ay,Luscher:1987ek,Luscher:1988uq,Huang:1988hu,Frick:1989gw,Kribs:2007nz}. The single-component $\phi^4$ theory becomes equivalent to the Ising model in the infinite limit of the quartic coupling $\lambda=\infty$ so that numerical studies of the $4d$ Ising model have been performed as a nonperturbative test of the triviality, assuming the universality \cite{PhysRevB.22.4481,SanchezVelasco:1987ah,Kenna:1992np,Bittner:2002pk,Kenna:2004cm,PhysRevE.80.031104,Lundow:2010en}. \footnote{In the standard model, we need to consider the $\phi^4$ interaction as a part of a combined Higgs-Yukawa sector, whose nonperturbative aspects were investigated with the lattice simulations \cite{Lee:1989xq,Lee:1989mi}. Also there are some recent studies to discuss the triviality of $O(N)$ $\phi^4$ theory with the higher-loop beta function \cite{Shrock:2014zca,Shrock:2016hqn,Shrock:2017zuk}.} So far, no Monte Carlo calculation has confirmed the logarithmic correction to the mean-field exponents in the scaling behavior of the specific heat, which is expected from the perturbative renormalization group analysis \cite{PhysRevB.7.248}. Moreover, a detailed Monte Carlo study has found a serious finite-volume effect due to nontrivial boundary effects in the $4d$ Ising model \cite{Lundow:2010en}.    

Recently, the authors have investigated the phase transition of the $4d$ Ising model with the higher-order tensor renormalization group (HOTRG) algorithm \cite{Akiyama:2019xzy}.
The tensor renormalization group (TRG) method,\footnote{In this paper, the TRG method or the TRG approach refers to not only the original numerical algorithm proposed by Levin and Nave \cite{Levin:2006jai} but also its extensions \cite{PhysRevB.86.045139,Adachi:2019paf,Kadoh:2019kqk,Shimizu:2014uva,Sakai:2017jwp,Akiyama:2020soe}.} which contains the HOTRG algorithm, has several superior features over the Monte Carlo method. (i) Since the TRG provides a deterministic numerical method, it does not have the sign problem encountered in stochastic methods, including the standard Monte Carlo simulation, as confirmed in various studies of quantum field theories \cite{Shimizu:2014uva,Shimizu:2014fsa,Shimizu:2017onf,Takeda:2014vwa,Kadoh:2018hqq,Kadoh:2019ube,Kuramashi:2019cgs,Akiyama:2020ntf}. (ii) Its computational cost depends on the system size only logarithmically. (iii) The computational cost to simulate fermions is almost equivalent to that to bosons because the TRG can directly manipulate the Grassmann variables \cite{Shimizu:2014uva,Sakai:2017jwp,Yoshimura:2017jpk,Akiyama:2020soe}. (iv) We can obtain the partition function or the path-integral itself. Thanks to the above feature (ii), we have been allowed to enlarge the lattice volume up to $V=1024^4$, which is essentially identified as the thermodynamic limit, and found finite jumps for the internal energy and the magnetization as functions of temperature in the $4d$ Ising model \cite{Akiyama:2019xzy}. These are characteristic features of the first-order phase transition. Having shown that the $4d$ Ising model undergoes the weak first-order phase transition, our interest turns to the order of the phase transition in the $4d$ single-component $\phi^4$ theory, which has the global $\mathbb{Z}_2$ symmetry as with the Ising model. \footnote{The scenario of the weak first-order phase transition in the Ising model or the $\phi^4$ theory has been discussed phenomenologically in some recent studies \cite{Cea:2019xdi,Consoli:2020nwb,Consoli:2020kip,Consoli:2020fer}.}

In this paper, we investigate the phase transition of the $4d$ single-component $\phi^4$ theory with the quartic coupling $\lambda$ and the hopping parameter $\kappa$, employing the anisotropic TRG (ATRG) algorithm \cite{Adachi:2019paf}, which was proposed to reduce the computational cost of the TRG method. The ATRG has been successfully applied to analyze the $4d$ complex $\phi^4$ theory at the finite density with parallel computation \cite{Akiyama:2020ntf}.
Our main purpose is to determine the order of the phase transition by examining the $\kappa$ dependence of the bond energy and the vacuum condensation of the scalar field $\braket{\phi}$ around the critical value of $\kappa_{\rm c}$ for the fixed $\lambda$, the latter of which is an order parameter of the phase transition caused by the spontaneous $\mathbb{Z}_2$ symmetry breaking.
We study the model with a single choice of $\lambda=40$, which is a finite-$\lambda$ generalization of the Ising model study performed in Ref.~\cite{Akiyama:2019xzy}, corresponding to $\lambda=\infty$. The choice of $\lambda=40$ may also be helpful to avoid the weak coupling region affected by the Gaussian fixed point at $\lambda=0$. For comparison, we also make the same analysis of the $3d$ single-component $\phi^4$ theory at $\lambda=40$, which is believed to belong to the universality class of the 3$d$ Ising model. We discuss the differences between the results of the 3$d$ and 4$d$ cases.  
  
This paper is organized as follows. In Sec.~\ref{sec:method} we explain the formulation of the lattice $\phi^4$ theory and the ATRG algorithm. We present numerical results for the 4$d$ and 3$d$ cases in Sec.~\ref{sec:results} and discuss the properties of the phase transition.  Section~\ref{sec:summary} is devoted to summary and outlook.

\section{Formulation and numerical  algorithm}
\label{sec:method}

We use the following popular action for the $d$-dimensional single-component $\phi^4$ theory on a lattice $\Gamma$:
\begin{widetext}
\begin{align}
	S[\phi]
	= \sum_{n\in\Gamma} \left[
		-\kappa \sum_{\nu=1}^{d} \left( \phi_n\phi_{n+\hat\nu}+\phi_n\phi_{n-\hat\nu} \right)
		+\phi_n^2 +\lambda\left( \phi_n^2 -1 \right)^2
	        \right],
        \label{eq:action_pop}
\end{align}
\end{widetext}
where $\hat\nu$ is the unit vector of the $\nu$-direction.
This formulation, which is explicit about the relation to the Ising model,
is equivalent to the more conventional expression
\begin{align}
	S[\varphi]
	=\sum_{n\in\Gamma} \left[
		\frac{1}{2}\sum_{\nu=1}^{d} \left( \varphi_{n+\hat\nu}-\varphi_n \right)^2
		+\frac{1}{2}m_0^2\varphi_n^2 +\frac{g_0}{4!}\varphi_n^4
	\right]
\end{align}
with
\begin{gather}
	\varphi_n=\sqrt{2\kappa}\phi_n, \\
	m_0^2=\frac{1-2\lambda}{\kappa}-2d, \\
	g_0=\frac{6\lambda}{\kappa^2}.
\end{gather}
The partition function is defined by
\begin{align}
	Z=\int\mathcal D\phi~\mathrm{e}^{-S[\phi]}
\end{align}
using the action of Eq.~\eqref{eq:action_pop} with the path integral measure
\begin{align}
	\int\mathcal D\phi=\prod_{n\in\Gamma}\int_{-\infty}^{\infty}\mathrm{d}\phi_n.
\end{align}
We express the partition function as a tensor network in the similar way to Ref.~\cite{Akiyama:2020ntf}.
The continuous variables $\phi_n$ are discretized by the $K$-point Gauss-Hermite quadrature rule as
\begin{align}
	\int_{-\infty}^\infty\mathrm{d}\phi_n~\mathrm{e}^{-\phi_n^2}f(\phi_{n})
	\simeq \sum_{\alpha_n=1}^K \omega_{\alpha_n} f(\phi_{\alpha_n}),
\end{align}
where $\phi_{\alpha}$ and $\omega_\alpha$ are the $\alpha$-th node and its weight.
The partition function is thus discretized as
\begin{align}
	Z(K) =\sum_{\{\alpha\}} \prod_{n,\nu} M_{\alpha_n \alpha_{n+\hat\nu}},
\end{align}
where
\begin{widetext}
\begin{align}
	M_{\alpha_n \alpha_{n+\hat\nu}}
	=\sqrt[2d]{\omega_{\alpha_{n}}\omega_{\alpha_{n+\hat{\nu}}}}\exp\left[
		2\kappa\phi_{\alpha_n}\phi_{\alpha_{n+\nu}}
		-\frac{\lambda}{2d}\left( \phi_{\alpha_n}^2-1 \right)^2
		-\frac{\lambda}{2d}\left( \phi_{\alpha_{n+\hat\nu}}^2-1 \right)^2
	\right].
\end{align}
\end{widetext}
Each matrix $M$ is approximated by the singular value decomposition (SVD) with a bond dimension $D$ as
\begin{align}
	M_{\alpha\beta}
	\simeq \sum_{k=1}^D U_{\alpha k} \sigma_k V_{\beta k},
\end{align}
where $\sigma_k$ is the $k$-th singular value sorted in the descending order,
and $U,V$ are the orthogonal matrices composed of the singular vectors. One finally obtains a tensor network representation for $Z(K)$ as
\begin{align}
\label{eq:part_tn}
	Z(K)
	=\sum_{\{i_{1},\cdots,i_{d}\}}\prod_{n\in\Gamma}
	T_{n;i_{1}\cdots i_{d}i'_{1}\cdots i'_{d}}, 
\end{align}
where
\begin{align}
\label{eq:initial}
	T_{n;i_{1}\cdots i_{d}i'_{1}\cdots i'_{d}}
	=\sum_{\alpha=1}^K \prod_{\nu=1}^{d}
	\sqrt{\sigma_{i_\nu}\sigma_{i'_\nu}} U_{\alpha i_\nu} V_{\alpha i'_\nu},
\end{align}
with the shorthand notations such as $i_{\nu}=i_{\nu,n}$ and $i'_{\nu}=i_{\nu,n-\hat{\nu}}$.

In this study, we employ the parallelized $d$-dimensional ATRG algorithm developed in Refs.~\cite{Akiyama:2020Dm,Akiyama:2020ntf}. We keep the bond dimension $D$ fixed throughout the ATRG procedure. 
For the swapping bond parts explained in Refs.~\cite{Adachi:2019paf,Oba:2019csk}, the randomized SVD is applied with the choice of $p=2D$ and $q=2D$, where $p$ is the oversampling parameter and $q$ is the numbers of QR decomposition. 

\section{Numerical results} 
\label{sec:results}

\subsection{4$d$ case}

The partition function of Eq.~\eqref{eq:part_tn} is evaluated using the ATRG algorithm on lattices with the volume $V=L^4$ ($L=2^m, m \in \mathbb{N}$) employing the periodic boundary conditions for all the space-time directions.
As explained in the previous section, there are two important algorithmic parameters. One is the number of nodes $K$ in the Gauss-Hermite quadrature method to discretize the scalar field. The other is the bond dimension $D$. We check the convergence behavior of the free energy as a function of $K$ and $D$ by defining the following quantities:
\begin{align}
	\delta_K=\left|\frac{\ln Z(K,D=50)-\ln Z(K=2000,D=50)}{\ln Z(K=2000,D=50)}\right|
\label{eq:delta_K}
\end{align}
and
\begin{align}
  \delta_D=\left|\frac{\ln Z(K=2000,D)-\ln Z(K=2000,D=50)}{\ln Z(K=2000,D=50)}\right|.
\label{eq:delta_D}
\end{align}

\begin{figure}[htbp]
	\centering
	\includegraphics[width=0.7\hsize]{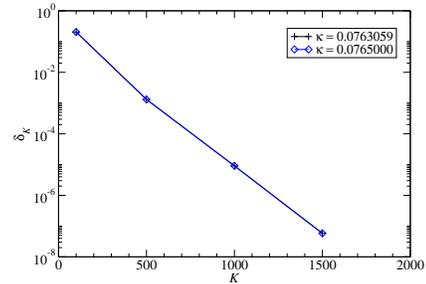}
	\caption{Convergence behavior of free energy with $\delta_K$ of Eq.~(\ref{eq:delta_K}) at $\kappa=0.0763059$ and $0.0765000$ as a function of $K$ on $V=1024^4$.}
	\label{fig:K_vs_F}
\end{figure}

\begin{figure}[htbp]
  	\centering
	\includegraphics[width=0.7\hsize]{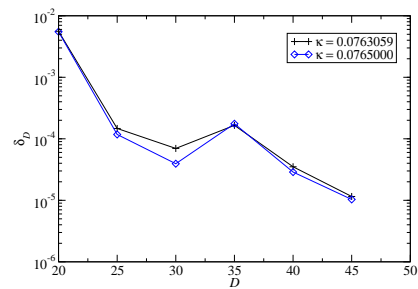}
  	\caption{Same as Fig.~\ref{fig:K_vs_F} for $\delta_D$ of Eq.~(\ref{eq:delta_D}).}
  	\label{fig:D_vs_F}
\end{figure}

Figure~\ref{fig:K_vs_F} shows the $K$ dependence of $\delta_K$ with $D=50$ on $V=1024^4$ at $\kappa=0.0763059$ and $0.0765000$, which are in the symmetric and broken symmetry phases. Note that $\kappa=0.0763059$ is close to the transition point $\kappa_{\rm c}$, as we will see below. We observe that $\delta_K$ decreases monotonically as a function of $K$ and reaches the order of $10^{-7}$ around $K=1500$. This shows that the Gauss-Hermite quadrature method is not affected by whether the system is in the symmetric or broken symmetry phase. We also plot the $D$ dependence of $\delta_D$ in Fig.~\ref{fig:D_vs_F}, which shows the fluctuation of free energy is suppressed as $\delta_{D}\approx10^{-5}$ up to $D=50$. Since the double-well potential in the $\phi^4$ theory becomes sharper for larger $\lambda$, we take a large number of $K$ to achieve good convergence for $\delta_K$. In the following, numerical results at $\lambda=40$ are presented for $K=2000$ and $D=50$ which are large enough in this study.

The phase transition point $\kappa_{\rm c}$ is determined by following the method employed in the Ising case \cite{Akiyama:2019xzy}.
Suppose we have obtained a coarse-grained tensor $T^{(m)}_{i_{1}i_{2}i_{3}i_{4}i'_{1}i'_{2}i'_{3}i'_{4}}$ after the $m$ times of coarse-graining. Defining a $D\times D$ matrix as
\begin{align}
\label{eq:matA}
	A^{(m)}_{i_{4}i'_{4}}=\sum_{i_{1},i_{2},i_{3}}T^{(m)}_{i_{1}i_{2}i_{3}i_{4}i_{1}i_{2}i_{3}i'_{4}},
\end{align}
we calculate 
\begin{align}
\label{eq:x}
	X^{(m)}=\frac{\left({\rm Tr} A^{(m)}\right)^2}{{\rm Tr} \left(A^{(m)}\right)^2}.
\end{align}
This quantity, introduced in Ref.~\cite{PhysRevB.80.155131}, possibly counts the number of the largest singular value of $A^{(m)}$. Therefore, it is expected that $X^{(m)}=1$ holds for the symmetric phase and $X^{(m)}=2$ for the broken symmetry phase. We may distinguish both phases by observing the plateau of $X^{(m)}$ after sufficient coarse-graining iterations.

\begin{figure}[htbp]
  	\centering
	\includegraphics[width=0.75\hsize]{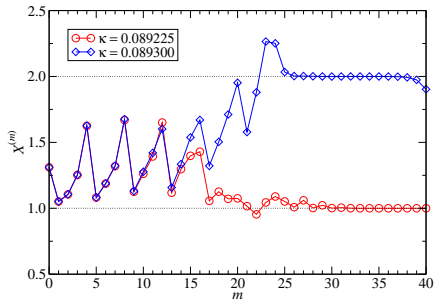}
	\caption{History of $X^{(m)}$ as a function of the coarse-graining step $m$ at $\kappa=0.089225$ (circle) and 0.089300 (diamond).}
  	\label{fig:x}
\end{figure}

\begin{figure}[htbp]
  	\centering
	\includegraphics[width=0.75\hsize]{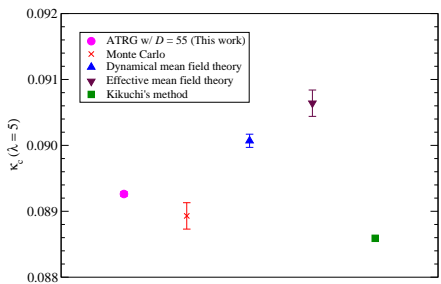}
	\caption{Comparison of $\kappa_{\rm c}$ at $\lambda=5$ obtained by various methods. All numerical values except for the ATRG result are taken from Table III in Ref.~\cite{Akerlund:2013fsa}. For details on the dynamical or effective mean field theory, see Ref.~\cite{Akerlund:2013fsa}. For Kikuchi's method, see Ref.~\cite{PhysRev.81.988}.}
  	\label{fig:lambda5}
\end{figure}

In order to check the applicability of the above method to determine the value of $\kappa_{\rm c}$, we calculate $\kappa_{\rm c}$ at $\lambda=5$ and compare it with the previous results obtained by various methods including the Monte Carlo simulation \cite{Akerlund:2013fsa}.
Since we have found that the convergence of the free energy with respect to the bond dimension at $\lambda=5$ becomes slightly slower than that at $\lambda=40$, we have taken $D=55$ (and $K=2000$) to evaluate $\kappa_{\rm c}$ at $\lambda=5$. Up to $D=55$, the relative error for the free energy is suppressed to $O(10^{-5})$.
Figure \ref{fig:x} shows the $m$ dependence of the value of $X^{(m)}$ at $\kappa=0.089225$ and $0.089300$, whose difference $\Delta \kappa=7.5\times10^{-5}$ is the finest resolution across the transition point. We find $X^{(m)}=1$ for $m\gtrsim 30$ at $\kappa=0.089225$ and $X^{(m)}=2$ for $m\gtrsim 25$ at $\kappa=0.089300$. Based on this observation, we determine the critical kappa $\kappa_{\rm c}=0.0892625(375)$ on the $1024^4$ lattice, whose error bar is provided by the resolution of $\kappa$. In Fig.~\ref{fig:lambda5} we find that our result is comparable to the Monte Carlo result $\kappa_{\rm c}=0.08893(20)$ in Ref.~\cite{Akerlund:2013fsa}. Slight deviation from the Monte Carlo result may be attributed to the finite size effect: our result is obtained on the $1024^4$ lattice, while the previous one is on the $32^4$ lattice. 

\begin{figure}[htbp]
  	\centering
	\includegraphics[width=0.75\hsize]{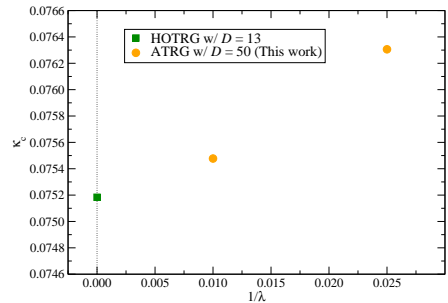}
	\caption{$\kappa_{\rm c}$ as a function of $1/\lambda$. $1/\lambda=0$ corresponds to the Ising model. Square symbol at $1/\lambda=0$ denotes the result obtained by the HOTRG~\cite{Akiyama:2019xzy}.
          All error bars are within symbols.}
  	\label{fig:kc}
\end{figure}

Having confirmed the validity of the method using $X^{(m)}$, we determine $\kappa_{\rm c}$ at $\lambda=40$ with $D=50$ and $K=2000$. The result is $\kappa_{\rm c}=0.076305975(25)$ on the $1024^4$ lattice, whose error bar is provided by the resolution of $\kappa$. In Fig.~\ref{fig:kc} we check the $1/\lambda$ dependence of $\kappa_{\rm c}$ toward the Ising limit, where the result at $\lambda=100$ is obtained in the same way as the $\lambda=40$ case with $D=50$ and $K=2000$. We observe that the value of $\kappa_{\rm c}$ seems monotonically approaching that in the Ising case. The error bars are provided by the resolution of $\kappa$ but they are all within symbols. 

\begin{figure}[htbp]
  	\centering
	\includegraphics[width=0.75\hsize]{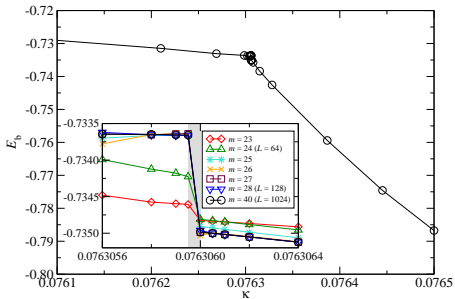}
 	 \caption{Bond energy as a function of $\kappa$ on $V=1024^4$. Inset graph shows it for various lattice sizes and gray band denotes $\kappa_{\rm c}$ estimated by $X^{(m)}$ of Eq.~\eqref{eq:x}.}
  	\label{fig:be}
\end{figure}

We now turn to the investigation of the phase transition with the bond energy defined by
\begin{align}
	E_{\rm{b}}=-\frac{1}{2}\frac{\partial}{\partial\kappa}\frac{\ln Z}{V}
\end{align}
and the vacuum condensation of the scalar field $\braket{\phi}$. Both quantities are evaluated with the impure tensor method. Figure~\ref{fig:be} plots the bond energy as a function of $\kappa$ on the $1024^4$ lattice. The resolution of $\kappa$ becomes finer toward the transition point and the finest one is $\Delta \kappa=5.0\times10^{-8}$ around the transition point. The phase transition point is consistent with $\kappa_{\rm c}$ (gray band) determined by $X^{(m)}$. Inset graph in Fig.~\ref{fig:be} shows an emergence of a finite gap with mutual crossings of curves for different volumes, $m\ge 23$, around $\kappa_{\rm c}$. These are characteristic features of the first-order phase transition as discussed in Ref.~\cite{Fukugita1990}. As the gap, we obtain
\begin{align}
	\Delta E_{\rm b}=0.001318(3),
\end{align}
by the linear extrapolation toward the transition point both from the symmetric and broken symmetry phases. In this extrapolation, we have used data points in $[0.07630560,0.07630595]$ for the symmetric phase and $[0.0763060,0.0763064]$ for the broken symmetry one. Note that we do not extrapolate $\Delta E_{\rm b}$ to the $D\rightarrow \infty$ limit in this paper because a systematic study of the $D$ dependence demands enormous computational cost and the theoretical formula for the extrapolation is not known so far. The value of $\Delta E_{\rm b}$ becomes smaller than the latent heat $\Delta E=0.0034(5)$ found in the Ising case with the HOTRG \cite{Akiyama:2019xzy}.

\begin{figure}[htbp]
	\centering
	\includegraphics[width=0.8\hsize]{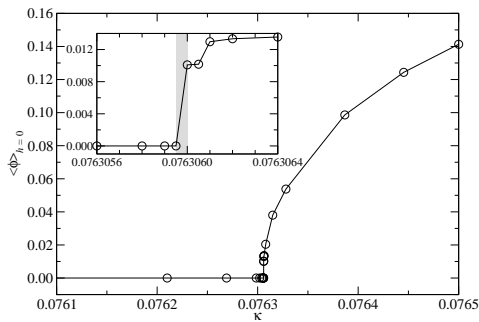}
 	 \caption{Vacuum condensation $\braket{\phi}_{h=0}$ as a function of $\kappa$ on $V=1024^4$. Gray band in inset graph shows $\kappa_{\rm c}$ estimated by $X^{(m)}$ of Eq.~\eqref{eq:x}.}
  	\label{fig:vc}
\end{figure}

Another quantity to detect the phase transition is the vacuum condensation of the scalar field $\braket{\phi}$, which is the order parameter of spontaneous breaking of the $\mathbb{Z}_2$ symmetry. We calculate $\braket{\phi}$ by introducing the external fields of $h=1.0\times 10^{-10}$ and $2.0\times 10^{-10}$ at each $\kappa$. After taking the infinite volume limit, we extrapolate the value of $\braket{\phi}$ at $h=0$. Figure~\ref{fig:vc} shows the $\kappa$ dependence of $\braket{\phi}_{h=0}$. The resolution of $\kappa$ is the same as that in Fig.~\ref{fig:be}. We find that the value of $\kappa_{\rm c}$, where the vacuum condensation sets in, is consistent with both estimates by $X^{(m)}$ and the bond energy.  A finite jump in $\braket{\phi}_{h=0}$ at $\kappa_{\rm c}$ is another indication of the first-order phase transition. We find
\begin{align}
	\Delta\braket{\phi}_{h=0}=0.0105(9), 
\end{align}
as the value of finite jump, where we have used data points in $[0.07630560,0.07630595]$ for the symmetric phase and $[0.0763060,0.0763064]$ for the broken symmetry one, as in the case with the bond energy, to extrapolate linearly the values of $\braket{\phi}_{h=0}$ toward the transition point. Note that this quantity is estimated as $0.037(2)$ in the Ising case with the HOTRG \cite{Akiyama:2019xzy}.

\subsection{3$d$ case}

The 2$d$ single-component lattice  $\phi^4$ theory is believed to belong to the same universality class as the 2$d$ Ising model. The previous TRG analysis, which was carried out by two of the authors and collaborators, supports this ansatz~\cite{Kadoh:2018tis}.
Although the 3$d$ case should undergo the second-order phase transition belonging to the universality class of the 3$d$ Ising model, the direct check with the TRG method has not been performed so far.  
Here it must be instructive to repeat the same TRG calculation for the 3$d$ case and compare the results between the 3$d$ and 4$d$ cases at $\lambda=40$.

We first show the convergence behavior of the free energy as a function of $K$ and $D$ by defining the relative error in the following way:
\begin{align}
	\delta_K=\left|\frac{\ln Z(K,D=90)-\ln Z(K=2000,D=90)}{\ln Z(K=2000,D=90)}\right|
\label{eq:delta_K_3d}
\end{align}
and
\begin{align}
  \delta_D=\left|\frac{\ln Z(K=2000,D)-\ln Z(K=2000,D=90)}{\ln Z(K=2000,D=90)}\right|.
\label{eq:delta_D_3d}
\end{align}

\begin{figure}[htbp]
	\centering
	\includegraphics[width=0.75\hsize]{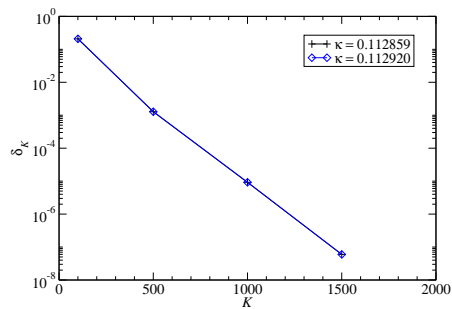}
 	\caption{Convergence behavior of 3$d$ free energy with $\delta_K$ of Eq.~(\ref{eq:delta_K_3d}) at $\kappa=0.112859$ and $0.112920$ as a function of $K$ on $V=4096^3$.}
  	\label{fig:K_vs_F_3d}
\end{figure}

\begin{figure}[htbp]
  	\centering
	\includegraphics[width=0.75\hsize]{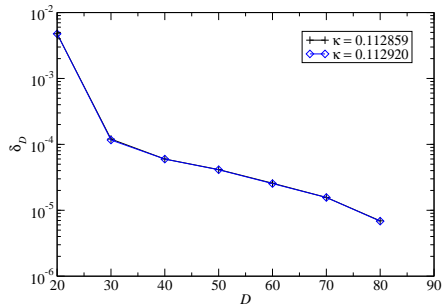}
  	\caption{Same as Fig.~\ref{fig:K_vs_F_3d} for $\delta_D$ of Eq.~\eqref{eq:delta_D_3d}.}
  	\label{fig:D_vs_F_3d}
\end{figure}

The $K$ dependence of $\delta_K$ with $D=90$ on $V=4096^3$ at $\kappa=0.112859$ and $0.112920$ in Fig.~\ref{fig:K_vs_F_3d}. $\kappa=0.112859$ is near the transition point in the symmetric phase, while $\kappa=0.112920$ is in the broken symmetric phase. 
We observe a monotonic decrease of $\delta_K$ as a function of $K$, which is quite similar to the 4$d$ case. Figure~\ref{fig:D_vs_F_3d} shows the $D$ dependence of $\delta_D$, where $\delta_D$ reaches the order of  $10^{-5}$ up to $D=90$. Notice that the achieved order of $\delta_D$ is similar with the $4d$ case. In the following, we present the results at $\lambda=40$ for $K=2000$ and $D=90$.

\begin{figure}[htbp]
  	\centering
	\includegraphics[width=0.75\hsize]{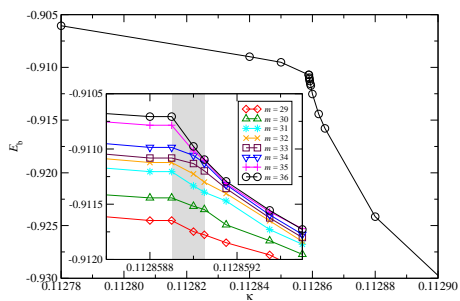}
 	 \caption{3$d$ bond energy as a function of $\kappa$ on $V=4096^3$ ($m=36$). Inset graph shows it for various lattice sizes and gray band restricts the location of $\kappa_{\rm c}$ by $X^{(m)}$.}
  	\label{fig:be_3d}
\end{figure}

Now let us discuss the results of the bond energy and the vacuum condensation of the scalar field $\braket{\phi}$, which are calculated with the impure tensor method as in the 4$d$ case. We plot the bond energy as a function of $\kappa$ on the $4096^3$ lattice in Fig.~\ref{fig:be_3d}, where the gray band with $0.11285890\le \kappa\le 0.11285905$ in the inset indicates the location of the phase transition point determined by $X^{(m)}$. Note that in the $3d$ case, $X^{(m)}$ is also given in the same way as Eq.~\eqref{eq:x}, defining the three-dimensional counterpart of Eq.~\eqref{eq:matA}.  The value of the bond energy evaluated at $\kappa=0.11285900$ is located within this gray band. This is due to the situation that $X^{(m)}$ at $\kappa=0.11285900$ does not show any clear plateau at $X^{(m)}=1$ or 2. We observe that the bond energy on all the volumes smoothly varies as a function of $\kappa$ without generating any gap. In addition, we find no mutual crossing of curves for different volumes around the phase transition point: The curve of the bond energy monotonically approaches that on the largest volume of $4096^3$. These behaviors, which are in clear contrast to the 4$d$ case, are characteristics of the second-order phase transition as discussed in Ref.~\cite{Fukugita1990}.

\begin{figure}[htbp]
	\centering
	\includegraphics[width=0.8\hsize]{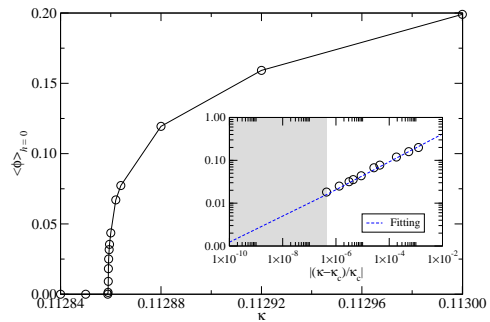}
	 \caption{3$d$ vacuum condensation $\braket{\phi}_{h=0}$ as a function of $\kappa$ on $V=4096^3$. Inset graph also shows $\braket{\phi}_{h=0}$ together with the fitting result (dotted line) as a function of the reduced parameter $|(\kappa-\kappa_{\rm c})/\kappa_{\rm c}|$ on a logarithmic scale. Gray band indicates $\kappa_{\rm c}$ estimated by $X^{(m)}$.}
 	\label{fig:vc_3d}
\end{figure}

In Fig.~\ref{fig:vc_3d}, we show the $\kappa$ dependence of $\braket{\phi}_{h=0}$, which is calculated in the same way as in the 4$d$ case. The resolution of $\kappa$ is the same as that in Fig.~\ref{fig:be_3d}. 
In order to determine the transition point $\kappa_{\rm c}$ and extract the critical exponent $\beta$, we make a fit of  $\braket{\phi}_{h=0}$ on $4096^3$ lattice, which is essentially in the thermodynamic limit, employing the function of $A(\kappa-\kappa_{\rm c})^\beta$ over the range of $\kappa \in[0.11285900,0.11300000]$ in the broken symmetry phase. The fit results are $A=3.7(9)$, $\kappa_{\rm c}=0.112859(6)$ and $\beta=0.32(2)$.
The value of $\beta$ is consistent with recent estimates of $\beta\approx 0.3295$ and 0.3264 for 3$d$ Ising model with the HOTRG algorithm~\cite{PhysRevB.86.045139} and the Monte Carlo method \cite{Hasenbusch:2011yya}, respectively. Numerical results for the bond energy and $\braket{\phi}_{h=0}$ show consistency with the second-order phase transition in the universality class of the 3$d$ Ising model.

\section{Summary and outlook} 
\label{sec:summary}

We have investigated the phase transition of the $4d$ single-component $\phi^4$ theory at $\lambda=40$ employing the bond energy and the vacuum condensation of the scalar field. Both quantities show finite jumps at the transition point on the extremely large lattice of $V=1024^4$, corresponding to the thermodynamic limit, and they indicate the weak first-order phase transition as found in the Ising limit~\cite{Akiyama:2019xzy}. This means that the single-component lattice $\phi^4$ theory does not have a continuum limit. In the current ATRG calculation, the resulting latent heat $\Delta E_{\rm b}$ and the gap $\Delta \braket{\phi}$ are smaller than those in the Ising case obtained by the HOTRG with $D=13$. As a next step, it would be interesting to investigate the phase transition of the O(4)-symmetric $\phi^4$ theory, which is more relevant to the SU(2) Higgs model. 

\begin{acknowledgments}
  Numerical calculation for the present work was carried out with the supercomputer Fugaku provided by RIKEN (Project ID: hp200170) and also with the Oakforest-PACS (OFP) and the Cygnus computers under the Interdisciplinary Computational Science Program of Center for Computational Sciences, University of Tsukuba.
This work is supported in part by Grants-in-Aid for Scientific Research from the Ministry of Education, Culture, Sports, Science and Technology (MEXT)
(No. 20H00148).
\end{acknowledgments}

\bibliography{formulation,algorithm,discrete,grassmann,continuous,gauge,review,for_this_paper}

\end{document}